\PassOptionsToPackage{sectionbib}{natbib}
\documentclass[journal=nalefd,manuscript=letter]{achemso}

\usepackage{amsmath, amsfonts,amssymb,color,graphicx,epsfig,float, textcomp}
\usepackage{caption}
\usepackage{xr}
\providecommand{\vect}[1]{\boldsymbol{#1}}

\graphicspath{{figs/}}

\title{Near-Perfect Absorption of Light by Plasmene Sheets}

\author{Qianqian Shi}
\affiliation{Department of Chemical Engineering, 
Faculty of Engineering, Monash University, Clayton, VIC 3800, Australia}

\author{Timothy U. Connell}
\affiliation{RMIT University, Melbourne, VIC, 3000, Australia}

\author{Qi Xiao}
\affiliation{CSIRO Manufacturing, Clayton, VIC, 3169, Australia}

\author{Anthony S. R. Chesman}
\affiliation{CSIRO Manufacturing, Clayton, VIC, 3169, Australia}
\affiliation{Melbourne Centre for Nanofabrication, Clayton, VIC, 3169, Australia }
\author{Wenlong Cheng}
\affiliation{Department of Chemical Engineering, 
Faculty of Engineering, Monash University, Clayton, VIC 3800, Australia}

\author{Ann Roberts}
\author{Timothy J. Davis}
\affiliation{School of Physics, The University of Melbourne, Parkville, VIC, 3010, Australia}

\author{Daniel E. G\'omez}
\email{daniel.gomez@rmit.edu.au}
\affiliation{RMIT University, Melbourne, VIC, 3000, Australia}


\begin{document}
	
\begin{abstract}

Near-perfect absorbers (NPAs) efficiently absorb visible light with a layered nanostructure that is thinner than the diffusion lengths of photogenerated charge carriers. 
We overcame existing limitations in fabricating their nanoparticulate surface by depositing \textit{plasmene}, a tightly-packed two-dimensional lattice of metal nanoparticles formed through self-assembly. 
The plasmene NPAs absorb up to 98\% of incident visible light, with modelling showing the improvement on existing NPAs arises from the structural ordering of the plasmene. 
We also demonstrate control of NPAs' absorption profile through the use of anisotropic building blocks in plasmene. These property enhancements may broaden the application of NPAs to structural colour, sensing and photocatalysis.
\end{abstract}

\section{Introduction}

Solar energy harvesting in photovoltaics and photocatalysis begins with the absorption of light by matter~\cite{Yu_POTNAOS2010a}.
Light absorption creates electronically excited states from which useful work can be extracted if the excited charge carriers can readily access chemical (catalytic site) or physical (electrode) interfaces.
To maximize solar-energy conversion efficiencies, it is thus desirable to create  light-harvesting devices in which the thickness of the absorbing layer is much smaller than characteristic charge-carrier scattering lengths, whilst simultaneously preserving a high magnitude of light absorption~\cite{Polman_NM2012a}.
\begin{figure}[hbt!]
\centering
\includegraphics[width=80mm]{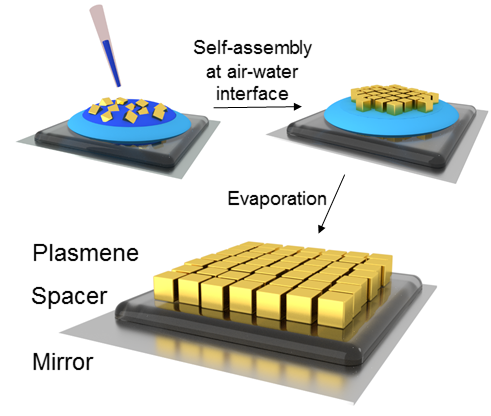}
\caption{The self-assembly of plasmene sheets on mirror/dielectric substrates (supported on glass), resulting in the formation of near-perfect absorbers.}
\label{fig:plasmene_npa}
\end{figure}

One class of structure that embodies these properties is the near-perfect absorber (NPA), in which near-complete light absorption occurs on a sub-wavelength scale due to optical impedance matching between the material and free space.
Optical impedance matching can be achieved by placing a layer of  nanoscale resonators  at a specific distance from a reflector, leading to the excitation of optical resonances that efficiently suppress the reflection  of light.
The presence of the mirror eliminates transmission, and consequently nearly all light is absorbed.
This concept has been demonstrated using metallic (plasmonic) nanostructures, 
resulting in strong absorption bands in the infrared~\cite{Liu_NL2010a, Chen_AN2012b,Bossard_AN2014a,Rozin_NC2015a} and visible~\cite{Hagglund_NL2013a,Akselrod_AM2015a} range of the electromagnetic spectrum.
Notably, there are remarkable examples where broadband light absorption has been achieved~\cite{Hedayati_AM2011a,Aydin_NC2011a,Liu_AAMI2015a,Ng_AN0a,Bossard_AN2014a}.
%
%
%

In most demonstrations of  plasmonic NPAs, the metallic nanostructures have been lithographically defined into periodic arrays or consist of randomly distributed nanoparticles, which result from the vacuum deposition and subsequent annealing of thin metal layers.
In contrast, creating these structures using colloidal suspensions of high quality nanoparticles would enable the facile fabrication of NPAs using simple and scalable deposition processes.
A theoretical study suggested that for a planar periodic array of lossy small particles (such as metal colloids) to exhibit  perfect absorption,  the absorption cross-section of an individual member of said array must be comparable to the area of the unit cell.~\cite{Thongrattanasiri_PRL2012a} 
This implies that the use of metal colloids as the building blocks of NPAs requires their assembly into surfaces of moderate to high particle densities.
To date, colloid-based plasmonic NPA structures have been realized by the deposition of  nanocubes using a self-assembly approach driven by the electrostatic attraction of colloids to the supporting substrates.
However, this leads to relatively low surface coverages ($\sim$ 4.2\% coverage~\cite{Moreau_N2012a}) and parasitic light scattering from the formed surfaces~\cite{Moreau_N2012a, Lassiter_NL2013a, Akselrod_AM2015a}.

Plasmene is  a highly ordered and tightly-packed two-dimensional sheet of metal nanoparticles formed by  spontaneous self-assembly at an air-water interface~\cite{Ng_AN2012a,Guo_TJOPCC2014a,Si_AN2014a,Si_AC2015a,Shi_AN2018a}.
These sheets are of single-particle thickness and form over macroscopic (i.e. $\sim$3 mm) lateral dimensions.
They also demonstrate several unique fundamental properties, such as an unusually high mechanical strength ($\sim$ 1GPa Young's modulus) and the capacity to support both propagating and gap plasmon modes, controllable by tuning the geometry of plasmene's unit cell~\cite{Si_AN2014a}.

Here, we demonstrate  near-perfect absorbers made with a nanostructured absorber layer of plasmene. The high surface coverage provided by plasmene creates NPAs that absorb up to 98\% of incident visible light. 
Furthermore, engineering the geometry of the plasmene building blocks allows for control of the wavelength of absorption.

\section{Results and Discussion}
\begin{figure}[htpb!]
\centering
\includegraphics[width=80mm]{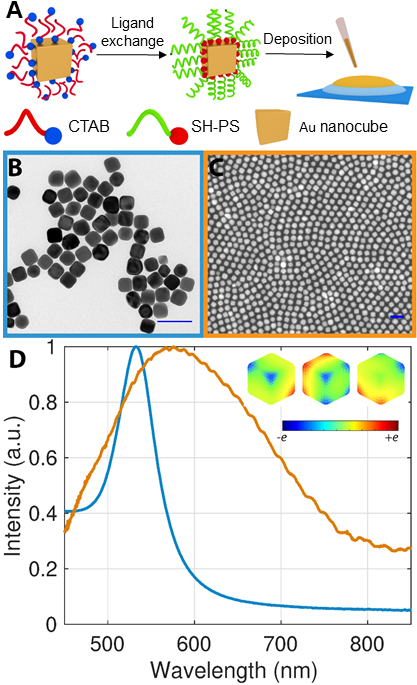}
\caption{ 
\textbf{A} The drying-mediated self-assembly of plasmene.
Electron microscope images of 
\textbf{B} Au nanocubes (side length $\sim$ 40 nm), and 
\textbf{C} a plasmene sheet.
The scale bar corresponds to 100 nm in both images. 
\textbf{D} Normalized absorption spectra of  Au nanocubes in solution and the extinction (1-T) spectrum of a Au nanocube plasmene sheet on glass. 
\textbf{Inset.} Calculated lowest-order degenerate  eigenmodes $\sigma_m(\vect r)$ of Au nanocubes.}
\label{fig:sem}
\end{figure}

The basic architecture of the plasmene NPAs, shown in Figure~\ref{fig:plasmene_npa}, consists of three distinct layers: a mirror, a thin dielectric spacer, and a single layer of plasmene.
Plasmene sheets with building blocks of Au nanocubes (shown in Figure~\ref{fig:sem}B) were synthesized following well-established protocols (Figure 2A)~\cite{Shi_AN2016a,Shi_AN2018a}.
Suspensions of these nanocrystals in water exhibit extinction bands in the visible that are characteristic of the excitation of localized surface plasmon resonances (Figure~\ref{fig:sem}D)~\cite{Davis_ROMP2016a}.
%
%

In solutions of low particle concentration, the inter-particle distances between neighboring particles is much larger than the near-fields generated by individual particle plasmons.
Consequently, the optical properties of these particles are well-described by considering the interaction of an isolated nanocube with an incident light field.
This interaction  excites localized surface plasmon resonances, creating a surface charge distribution $\sigma(\vect r)$.
Given that the nanoparticle dimensions are significantly smaller than the wavelength of visible light, this charge distribution can be written in terms of the (electrostatic) {\it eigenmodes} $\sigma_m(\vect r)$ (or self-sustained oscillations) of the particle according to  
$\sigma(\vect r) = \sum_m a_m(\omega)\sigma_m(\vect r)$, 
where   $a_m(\omega)$ are the \textit{excitation amplitudes} that describe the coupling of the incident light field with a particular eigenmode (here $m$ is an index that indicates the $m$-th mode sustained by the nanoparticle).
Modes with strong dipolar character couple strongly with uniform electric fields and thus have large values of $a_m(\omega)$. 
Consequently, these are manifest in  optical extinction spectra, as shown in Figure~\ref{fig:sem}D corresponding to Au nanocubes in Figure~\ref{fig:sem}B.
The calculated lowest-order (three-fold) degenerate eigenmodes $\sigma_m(\vect r)$ of the nanocubes clearly have a strong dipolar character, with charge accumulation at each of the three orthogonal directions of the cube (Figure~\ref{fig:sem}D inset).
For these particles, the electrostatic eigenmode description (particles in a uniform medium with the refractive index of water; eigenvalue for this mode 1.728) predicts that these dipole  resonances occur at 560 nm,  which is in good agreement with the experimental value. 

Plasmene is formed by the deposition of a chloroform solution of Au nanoparticles onto the surface of a water droplet. A change in polarity resulting from the evaporation of the chloroform results in self-assembly into an extended two-dimensional array on the surface of the water droplet, with subsequent evaporation yielding a quasi-crystalline lattice of plasmene~\cite{Bauer_PRB2011a} (Figure~\ref{fig:sem}A, more detail in the Methods Section).
The extinction spectrum ($1-T$) of a plasmene layer on a glass surface  clearly demonstrates a strong modification to the nanostructures' interaction with light (Figure~\ref{fig:sem}D).
This results from a combination of at least two effects: 
the interaction of the nanoparticles with a flat substrate, as well as inter-nanoparticle interactions.
When the nanocubes are placed on a substrate, the degeneracy of the dipolar modes shown in Figure~\ref{fig:sem}D is lifted, leading to hybridisation between bright (dipolar) and dark (multipolar) plasmon modes.
In the case of Ag nanocubes, it has been shown that this results in substantial spectral red-shifts  and the emergence of resonances that were absent in the solution spectrum~\cite{Sherry_NL2005a,Zhang_NL2011a}. 

Inter-nanoparticle interactions also lead to plasmon hybridization~\cite{Prodan_S2003a}; the surface electron charge oscillations $\sigma(\vect{r})$  
that result from the interaction with light are delocalized over the array of particles.
This forms a \textit{collective} plasmon mode with a resonance frequency red-shifted with respect to that of the isolated particle.
%
%
\begin{figure}[htpb!]
\centering
\includegraphics[width=80mm]{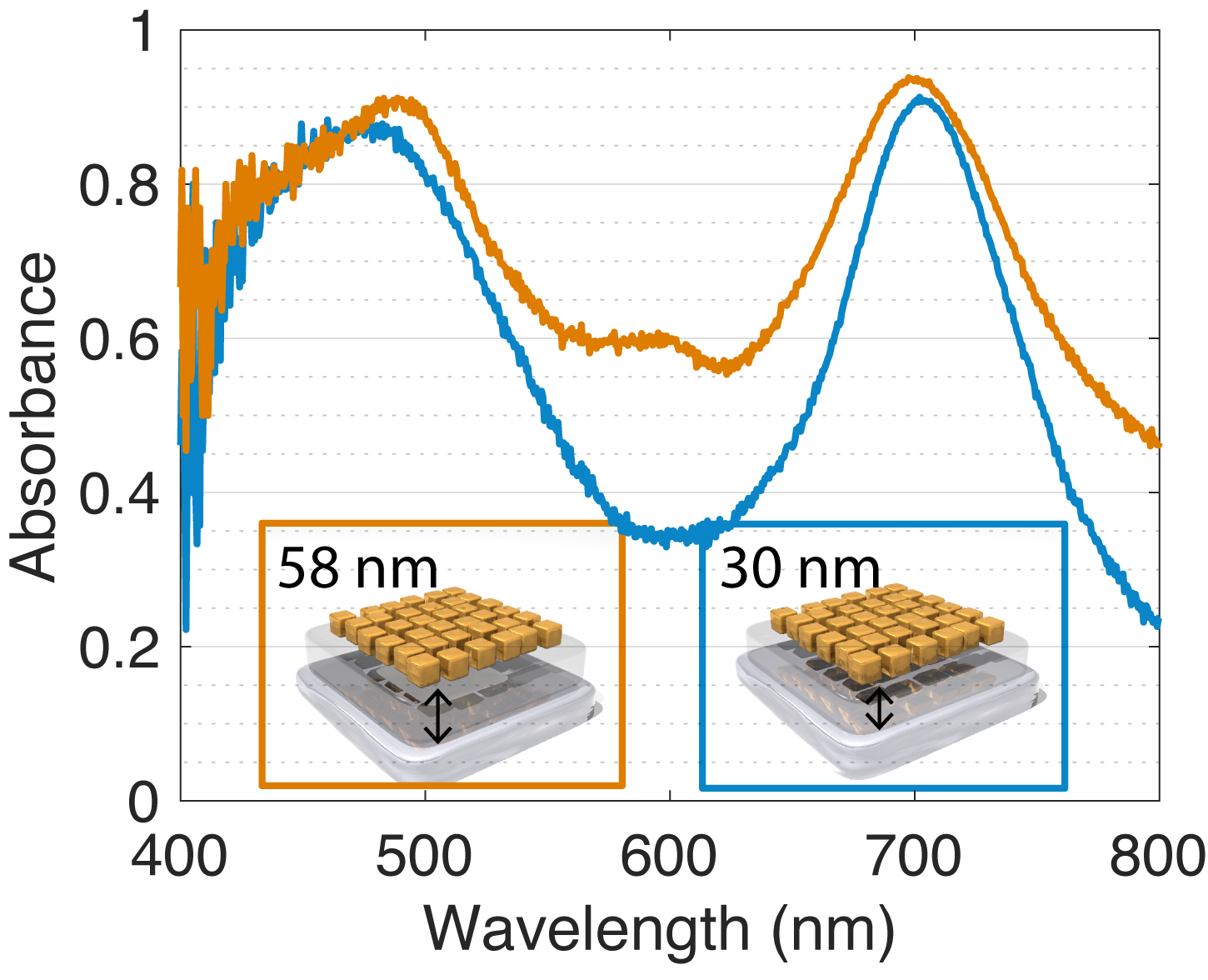}
\includegraphics[width=80mm]{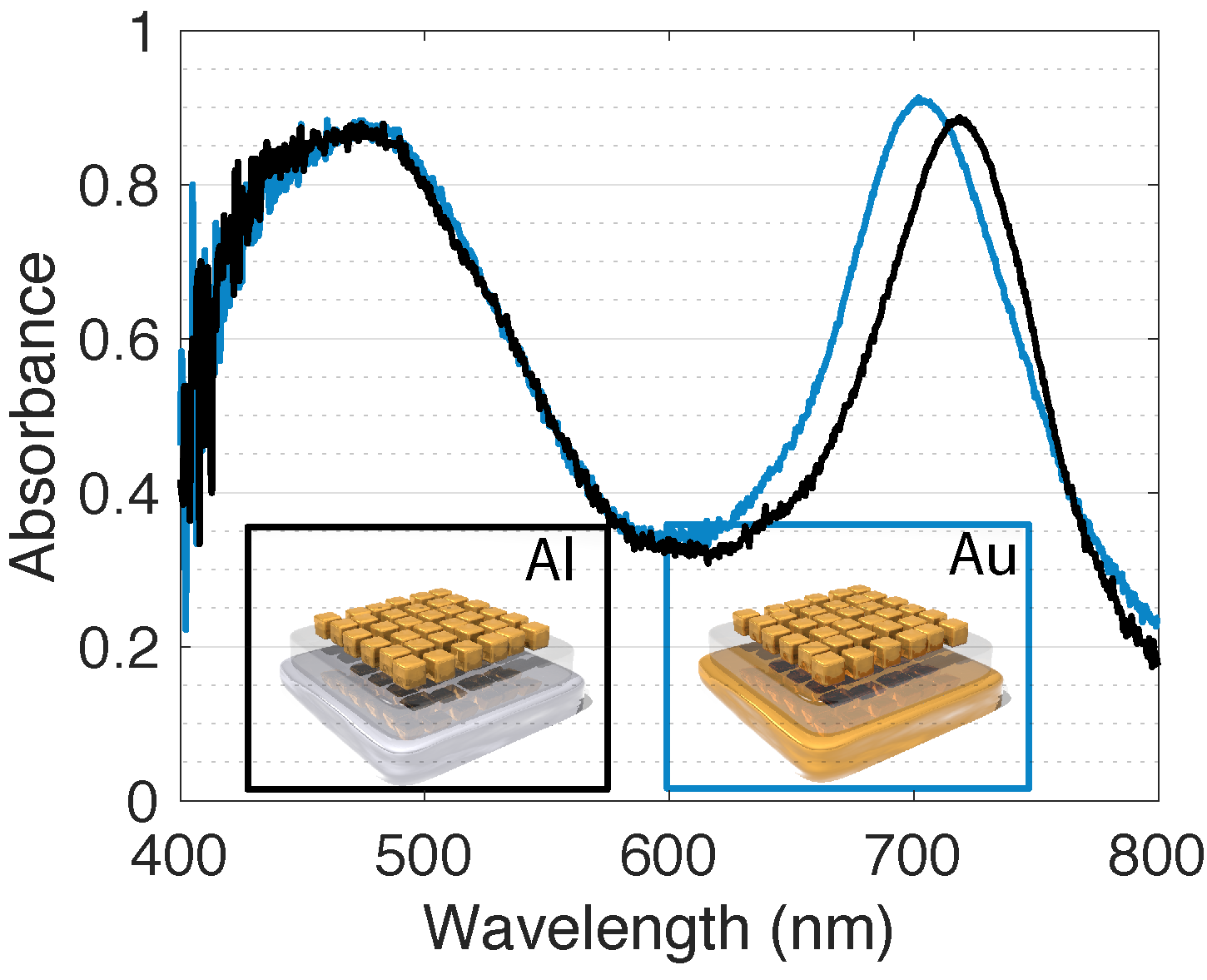}
\caption{\textbf{A} Absorption spectra of plasmene NPAs made with Au nanocubes / TiO$_2$ / Al mirror, with varying TiO$_2$ thicknesses affecting the absorption spectra. \textbf{B} Effect of mirror type on the absorption spectra of plasmene near-perfect absorbers (TiO$_2$ thickness fixed at 30 nm.)}
\label{fig:3}
\label{fig:mirror}
\end{figure}

To experimentally realize plasmene  near-perfect absorbers of light, the mirror and spacer layer were deposited by physical vapor deposition. 
The spacer layer consisted of TiO$_2$, a dielectric with a high refractive index that is transparent to visible light.
Self-assembly of a plasmene sheet on the TiO$_2$ thin film completed the structure.
Figure~\ref{fig:3}A shows the measured absorption spectra of two samples constructed using Al as the mirror, Au nanocubes as the plasmene building blocks, and TiO$_2$ of different thicknesses.
These spectra were determined by measuring the transmission ($T$) and reflection ($R$) spectra of each sample (the latter carefully referenced to a surface with a known reflectance spectrum), giving the absorption ($A$) as: 
$A = 1 - R - T$ by virtue of the conservation of energy (more detail in the Methods Section, the transmission spectra are shown in Figure \ref{fig:s2}).
With a dielectric spacer of 30 nm thickness, the spectrum showed two dominant absorption bands at
480 and 703 nm reaching absorption magnitudes of 86\% and 91\%, respectively.
Increasing the dielectric spacer thickness to 58 nm slightly improved the maximum absorption to  
94\% at 700 nm.
Interestingly, whilst the absorption band at $\sim$700 nm underwent a slight hypsochromic shift, the higher energy band ($\sim$480--500 nm) showed a bathochromic shift as the TiO$_2$ thickness increased, consistent with a distance-dependent interaction between the plasmene sheet and the metal layer~\cite{Lassiter_NL2013a}.
%
Changing the chemical composition of the mirror layer in the NPA from Al to Au caused a hypsochromic shift in the low energy absorption band (approx. 10 nm) along with a small increase in magnitude (Figure~\ref{fig:mirror}B). 
When compared to the extinction spectrum of a nanocube plasmene sheet deposited on glass (Figure~\ref{fig:sem}C), the absorption bands of Figures~\ref{fig:3} are noticeably sharper. 
%
\begin{figure}[htpb!]
\centering
\includegraphics[width=80mm]{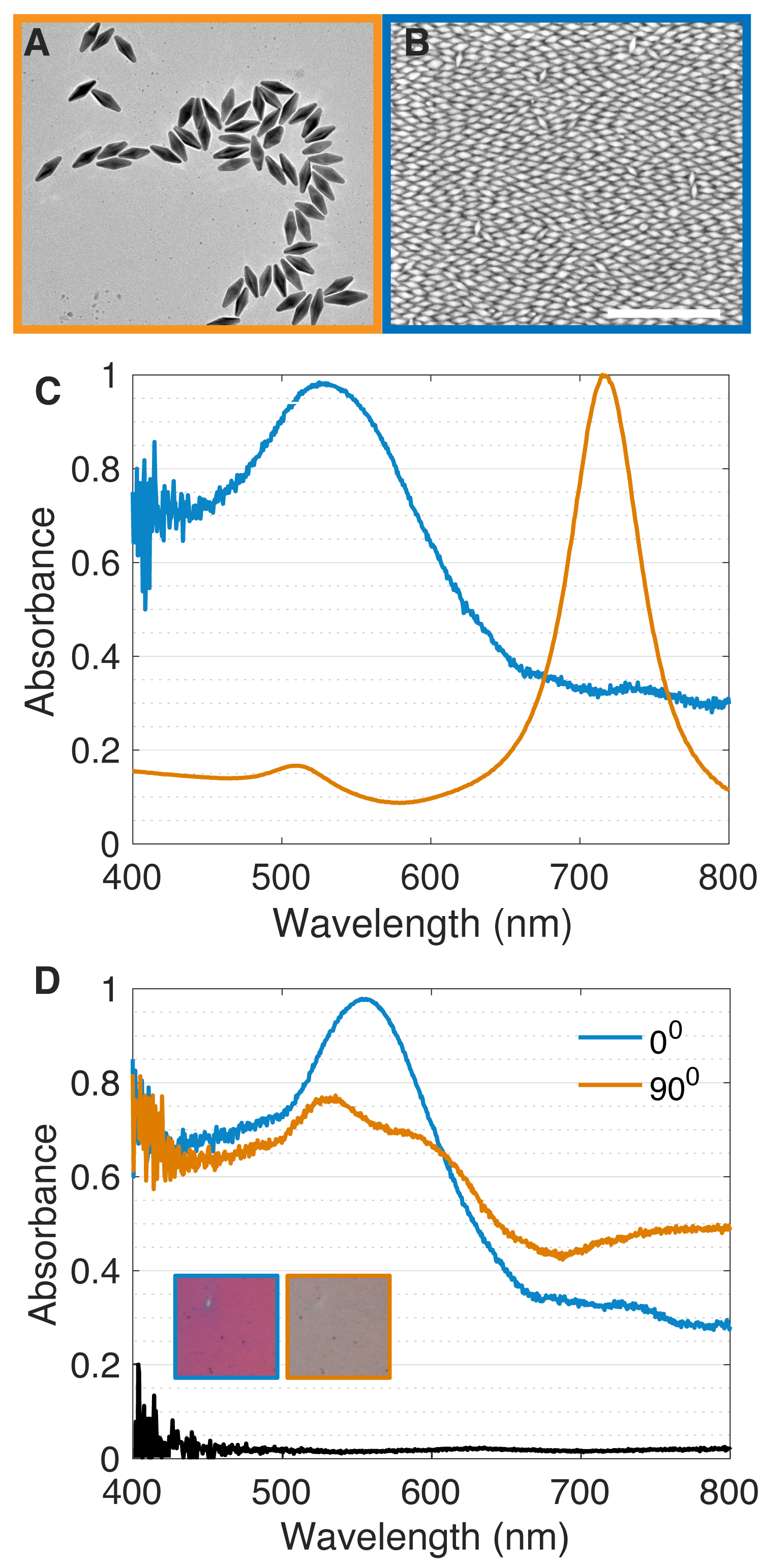}
\caption{Effect of plasmene building block geometry on optical properties.
Electron microscope images of 
\textbf{A} Au bipyramids in solution (scale bar 100 nm) and 
\textbf{B} a plasmene sheet made with Au bipyramids (scale bar 400 nm), illustrating the lattice orientation that is responsible for the observed polarization dependence.
\textbf{C} (Normalized) absorption spectrum of Au bipyramids in solution and absorption spectrum of a plasmene near-perfect absorber made with bipyramids.
\textbf{D} Absorption spectra for plasmene sheets made with Au bipyramids measured for two polarizations.
The black line corresponds to the measured dark-field scattering spectrum. 
For the cases shown here, the mirror was made with Al and the spacer consisted of 30 nm of TiO$_2$.
\textbf{Inset}. Optical microscope images of the measured regions demonstrating a drastic change in color.
}
\label{fig:DF}
\end{figure}

The absorption spectrum of plasmene is spectrally tailored by the geometry of the nanoparticle building blocks. 
This effect is illustrated in Figure~\ref{fig:DF}, where we show  NPAs made with bipyramidal-shaped Au nanocrystals. 
Due to their long aspect ratio (Figure~\ref{fig:DF}A), Au bipyramid  nanoparticles exhibit multiple localized surface plasmon resonances (Figure~\ref{fig:DF}C).
A plasmene NPA incorporating bipyramidal building blocks exhibited an absorption band with a maximum of 98\% incident light absorption at 524 nm.
(A second absorption band is expected to occur at higher wavelengths, beyond the detection range of the instrumentation used). 
Nanobipyramids can self-assemble to form plasmene sheets in four different packing orders~\cite{Shi_AN2016a}, and due to their high structural aspect ratio,  these assemblies can form plasmene sheets with optical properties that are sensitive to the polarization of the incident electromagnetic field (Figure~\ref{fig:DF}D).
These changes are also clearly visible to the naked eye under an optical microscope (Figure~\ref{fig:DF}D inset).

The scattering of light imposes a lower limit on the reflectivity at the peak absorption.
Scattering can occur as a consequence of structural disorder or the existence of edges in the plasmene lattice.
The former may arise from the size distribution of the building blocks, whilst the latter may occur due to the finite extent of the plasmene sheets or the unintentional formation of stacked multiple layers.
Figure~\ref{fig:DF}(D)  shows a comparison between the measured  absorption  and dark-field scattering spectra of a plasmene near-perfect absorber, which were measured against  surfaces of known reflectance and calibrated scattering amplitude, respectively.
The figure shows clearly that scattering amounts to $<$ 1\% of the total incident power, indicating that it is strongly suppressed due to the regular ordering of the nanocrystal building blocks in the plasmene layer~\cite{Moreau_N2012a}.
%

\textbf{Theory}
%
For metasurfaces with a nanostructured layer, in which the inter-particle spacings are larger than the spatial extent of the plasmonic near-fields, perfect absorption has been modelled with a  coupled-mode theory~\cite{Ciraci_JOAP2013a,Bowen_PRB2014a}, wherein  
each individual mirror-coupled nanocube reacts like a magnetic dipole.
The collective action of an ensemble of these nanocubes results in an effective magnetic response, which enables almost complete  impedance matching between the metasurface and free space, resulting in near-perfect absorption~\cite{Moreau_N2012a,Lassiter_NL2013a,Akselrod_AM2015a}.
In metasurfaces with a sparse surface coverage, these magnetic modes can be both excited and scattered at the edges of the nanocube, allowing for some of the radiation to escape and couple into free space.

\begin{figure}[htpb!]
	\centering
	\includegraphics[width=80mm]{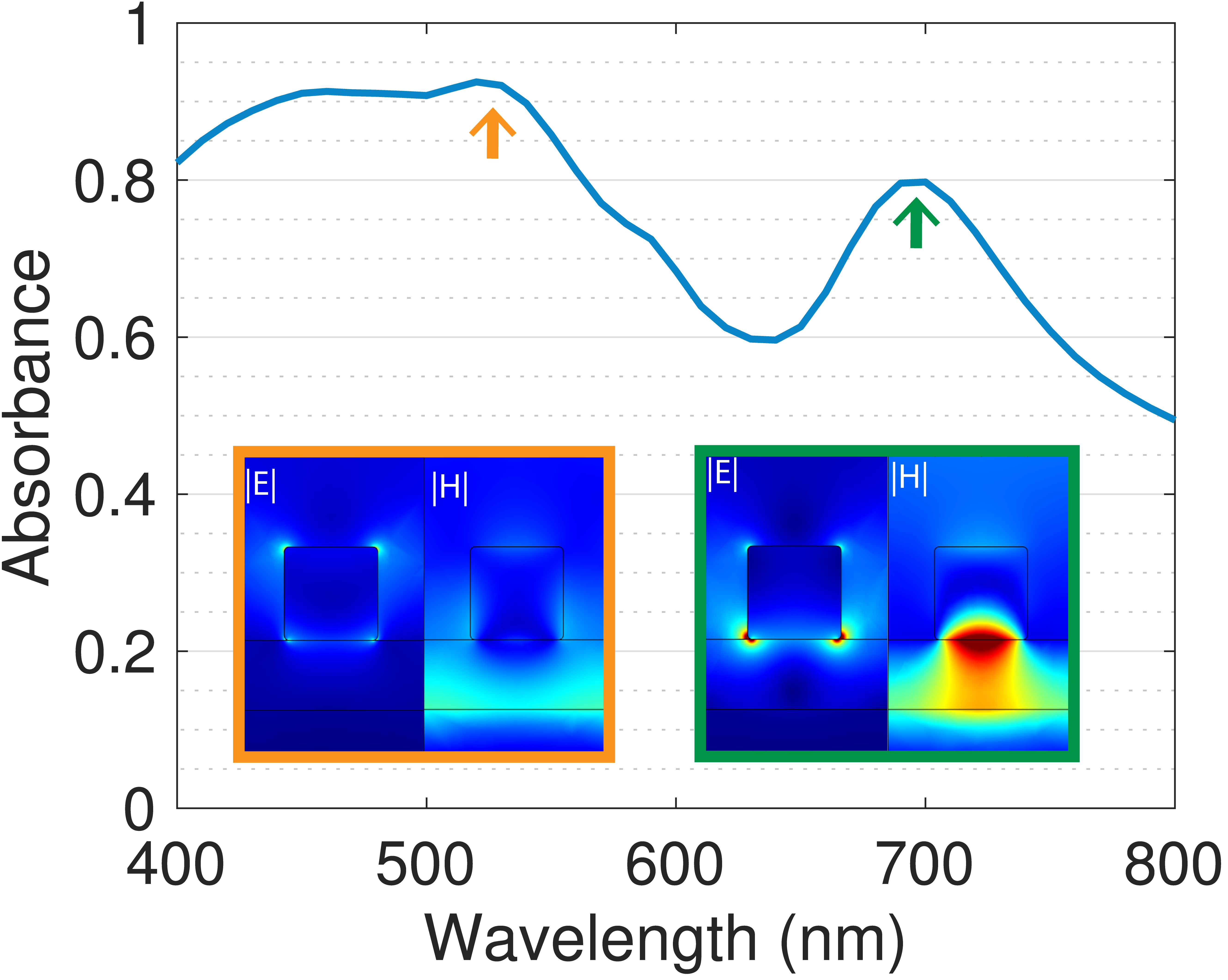}
	\caption{Numerically simulated optical absorption spectrum for a perfectly-ordered array of Au nanocubes on a TiO$_2$--Au mirror support. The insets show spatial maps of the magnitude of the electric ($|E|$) and magnetic ($|H|$) fields. The magnetic field distribution for the absorption peak located at $\sim$700 nm clearly shows that at this wavelength, normally-incident light excites a magnetic mode in the structure.}
	\label{fig:rcwa}
\end{figure}

In order to gain a physical insight into the mechanism that enables almost complete impedance matching in plasmene NPAs, we have modelled the optical properties of a nanocube plasmene NPA using a finite element approach for solving Maxwell equations.
The  model material consisted of a periodic array of nanocubes of side length 40 nm ordered in a cubic lattice with a 100 nm period and a 30 nm spacer layer.
Figure~\ref{fig:rcwa} shows the calculated absorption  spectrum for this configuration with two bands that closely match the experimental data, along with  the calculated maps of the magnitude of the electric ($|E|$) and magnetic ($|H|$) fields for each band.
The electric field  is strongly localized at the edges of the nanocube distant from the dielectric film, for the short--wavelength absorption band (located at $\sim$ 540 nm).
For the long-wavelength absorption band (located at $\sim$700 nm), the electric field is strongly localized  at the edges of the cube in contact with the dielectric film, and the magnetic field exhibits strong localization in the TiO$_2$ layer, in a way similar to a magnetic dipole~\cite{Liu_NL2010a}.
This optical magnetic response originates from the interaction of the induced oscillating surface charges on the cube with their mirror images on the metal support.
The radiation from the latter can cancel that produced by the induced dipole located on the nanoparticles, leading to suppression of reflection and causing the nanoparticles to absorb the energy, given that the mirror support prevents transmission of light~\cite{Chen_OE2012a}. 
Given the deep sub-wavelength separation between the building blocks, the plasmene sheet can be rationalized to consist of an effective \textit{magnetic metasurface}, that is different from a metal film of a similar thickness (see Figure~\ref{fig:s1}). 


\section{Conclusion}
In summary, we have fabricated and measured the absorption spectra of plasmene near-perfect absorbers of light.
We demonstrate that these materials can absorb up to 98\% of incident visible light.
The intense electromagnetic near-fields resulting from this near total extinction of light may find application in surface-enhanced spectroscopy, chemical and biological sensing, and plasmonic hot-carrier extraction. Notably, as the mirror and TiO$_2$ layers can be deposited by solution-based approaches (such as electroplating, spin-coating and sol-gel methods), this report provides a pathway for the scalable, vacuum-free fabrication of NPAs.


\section{Methods}
\textbf{Methods}

Require a general materials section. Also, should note that all solutions are aq. unless otherwise noted, this is ambiguous.

\textbf{Synthesis of Nanocrystal Building Blocks}
\textit{Synthesis of Au nanocubes}
A seed-mediated method was used to synthesize Au nanocubes~\cite{Qianqian_AMa}.. 
Briefly, a seed solution was prepared by adding NaBH$_4$ (0.6 ml, 0.01 M) to an aqueous solution prepared by mixing HAuCl$_4$ (0.1 ml, 25 mM) and cetyltrimethylammonium bromide (CTAB) (7.65 ml, 0.1 M). 
The seed solution was then kept at 30 $^\circ$C for 1 hour. 
The growth of Au nanocubes was triggered by adding 5 $\mu$l of a 10-fold diluted seed solution into a growth solution prepared by adding CTAB (1.6 ml, 0.1 M), HAuCl$_4$ (0.2 ml, 7 mM), and ascorbic acid (1.2 ml, 0.1 M) to Milli-Q water (8 ml). 
The resulting solution was then left undisturbed overnight, followed by centrifugation at 7830 rpm for 10 min and redispersal in Milli-Q water (10 ml) for further use.

\textit{Synthesis of Au Nanobipyramids (NBP)}
The synthesis of Au nanobipyramids involved a seed-meditated method and a three-step purification, as reported~\cite{Shi_AN2016a}. 
A Au-seed solution was synthesized by adding NaBH$_4$ (1 ml, 100 mM) to a solution prepared by adding HAuCl$ _4$ (0.4 ml, 25 mM) to aqueous trisodium citrate (40 ml, 0.25 mM). 
The resulting solution was then kept stirring at room temperature for 2 hours. 
Later, a portion of this seed solution (0.8 ml) was added to 100 ml of growth solution that contained aliquots of CTAB (0.1 M), HAuCl$_4$ (2 ml, 25 mM), AgNO$_3$ (1 ml, 10 mM), HCl (2 ml, 1.0 M), and ascorbic acid (0.8 ml, 0.1 M). 
The growth solution was kept at 30 \textdegree{}C overnight, and then centrifuged at 7830 rpm for 10 min, with the precipitate redispersed in an aqueous solution of CTAC (100ml, 80 mM). 
Purification of the Au NPBs was achieved using a three-step method reported previously~\cite{Shi_AN2016a}. 
The final solution was washed and redispersed in Milli-Q water for further use.

\textbf{Preparation of Metal-Semiconductor Substrates}
All films were deposited in an electron beam evaporation system (Intlvac, Nanochrome II) using a 10 kV power supply in a Class 10,000 cleanroom.
Glass substrates were cleaned by sequential ultra-sonication in acetone and isopropanol, and then dried under a stream of nitrogen gas.
The substrates were also cleaned by a plasma treatment step prior to the deposition of the metal-semiconductor films.
In order to achieve good adhesion of the mirror layer to the substrate, a thin layer of Cr (5 nm, deposited 0.5 \AA{}/s) was deposited. 
The Al or Au mirrors (100 nm, deposited 1.0 \AA{}/s), and TiO$_2$ spacer (30 or 60 nm, deposited 0.5 \AA{}/s) were deposited sequentially.
Dielectric film thickness was confirmed by spectroscopic ellipsometry (J. A. Woolam Co., M-2000DI).

\textbf{Fabricated of Plasmene Sheets}
Plasmene was fabricated by a drying-mediated self-assembly method at an air-water interface~\cite{Shi_AN2018a}. 
First, a two-step ligand-exchange method was used to replace CTAB ligands with thiol-terminated polystyrene (SH-PS) (Mn=50,000 for Au nanocubes and 20,000 for NBP). 
Briefly, 10 ml of the CTAB-capped Au nanocube solution (or 12 ml of the Au NBP solution) was centrifuged and redispersed in 5 ml (or 6 ml for Au NBP) SH-PS-THF (4 mg/ml) solution. 
The solution was then kept at room temperature overnight and purified by repeated centrifugation-precipitation cycles to remove uncapped ligands and precipitate the SH-PS capped nanoparticles. 
Finally, the solution was concentrated and redispersed in a small amount of chloroform.
To assemble a plasmene sheet, one drop of nanoparticle solution was deposited onto a water droplet sitting on the mirror/spacer substrate. 
After the full evaporation of the water and chloroform, a plasmene sheet was obtained.

\textbf{Characterization}
Normal incidence images of the resulting structures were obtained by scanning electron microscopy (FEI, NovaNanoSEM 430). 
SEM images of  plasmene were obtained by a FEI Helios Nanolab 600 FIB-SEM operating at 5 kV.  

\textbf{Optical Measurements} 
The optical transmission and reflectance spectra of the NPA plasmene sheets were measured using an upright  microscope (Nikon LV100), coupled to a spectrograph (Andor, SR-303i-A) equipped with a CCD (iDUS DU420A-BEX2-DD). 
For all the measured spectra, the light source was a broadband 100 W halogen lamp. 
The illumination and collection of reflectance ($R$) was done with a ${50}\times$ objective (NA 0.55), and these measurements were referenced to a mirror with a known reflectance spectrum (Thorlabs BB1-E02).
The transmission ($T$) spectra were calibrated using a bare glass substrate. 
The absorption data shown in this paper were calculated as: 
$A = 1 - R - T$ and are consequently not arbitrary values~\cite{Aydin_NC2011a}. 
Scattering spectra were measured using the same objective but placing a dark-field iris in front of the light source.
These spectra were referenced to a white diffuse Reflectance Standard (Labsphere, Spectralon SRS-99-020).

\textbf{Numerical Simulations}
The numerical simulations shown in Figure \ref{fig:rcwa} 
were undertaken using the Finite Element Method implemented in COMSOL Multiphysics 5.3a. 
We used tabulated data for the refractive index of Au~\cite{Johnson_PRB1972a} and Al~\cite{Rakic} for the substrate, and 1.0 for the medium above the structure. 
The refractive index of the TiO$_2$ layer was assumed to be 2.4. 
Periodic boundary conditions were assumed on the sides and a port boundary was used at the excitation surface. 
The normally incident plane was assumed to be linearly polarised parallel to one of the horzontal edges of the cubes. 
The other surface of the model adjacent to the Al was taken to be a scattering boundary. 
No apparent backscattering from the terminating boundaries was apparent. 
The cubes were rounded to have a radius of curvature of 2 nm.

\textbf{Acknowledgements}
This work was performed in part at the Melbourne Centre for Nanofabrication (MCN) in the Victorian Node of the Australian National Fabrication Facility (ANFF).
We acknowledge the ARC for support through a Future Fellowship (FT140100514) and a Discovery Project (DP160100983).

\textbf{Competing financial interests}
The authors declare no competing financial interests

\bibliography{references}

\providecommand{\latin}[1]{#1}
\makeatletter
\providecommand{\doi}
  {\begingroup\let\do\@makeother\dospecials
  \catcode`\{=1 \catcode`\}=2\doi@aux}
\providecommand{\doi@aux}[1]{\endgroup\texttt{#1}}
\makeatother
\providecommand*\mcitethebibliography{\thebibliography}
\csname @ifundefined\endcsname{endmcitethebibliography}
  {\let\endmcitethebibliography\endthebibliography}{}
\begin{mcitethebibliography}{33}
\providecommand*\natexlab[1]{#1}
\providecommand*\mciteSetBstSublistMode[1]{}
\providecommand*\mciteSetBstMaxWidthForm[2]{}
\providecommand*\mciteBstWouldAddEndPuncttrue
  {\def\EndOfBibitem{\unskip.}}
\providecommand*\mciteBstWouldAddEndPunctfalse
  {\let\EndOfBibitem\relax}
\providecommand*\mciteSetBstMidEndSepPunct[3]{}
\providecommand*\mciteSetBstSublistLabelBeginEnd[3]{}
\providecommand*\EndOfBibitem{}
\mciteSetBstSublistMode{f}
\mciteSetBstMaxWidthForm{subitem}{(\alph{mcitesubitemcount})}
\mciteSetBstSublistLabelBeginEnd
  {\mcitemaxwidthsubitemform\space}
  {\relax}
  {\relax}

\bibitem[Yu \latin{et~al.}(2010)Yu, Raman, and Fan]{Yu_POTNAOS2010a}
Yu,~Z.; Raman,~A.; Fan,~S. \emph{Proceedings of the National Academy of
  Sciences} \textbf{2010}, \emph{107}, 17491--17496\relax
\mciteBstWouldAddEndPuncttrue
\mciteSetBstMidEndSepPunct{\mcitedefaultmidpunct}
{\mcitedefaultendpunct}{\mcitedefaultseppunct}\relax
\EndOfBibitem
\bibitem[Polman and Atwater(2012)Polman, and Atwater]{Polman_NM2012a}
Polman,~A.; Atwater,~H.~A. \emph{Nat Mater} \textbf{2012}, \emph{11},
  174--177\relax
\mciteBstWouldAddEndPuncttrue
\mciteSetBstMidEndSepPunct{\mcitedefaultmidpunct}
{\mcitedefaultendpunct}{\mcitedefaultseppunct}\relax
\EndOfBibitem
\bibitem[Liu \latin{et~al.}(2010)Liu, Mesch, Weiss, Hentschel, and
  Giessen]{Liu_NL2010a}
Liu,~N.; Mesch,~M.; Weiss,~T.; Hentschel,~M.; Giessen,~H. \emph{Nano Letters}
  \textbf{2010}, \emph{10}, 2342--2348\relax
\mciteBstWouldAddEndPuncttrue
\mciteSetBstMidEndSepPunct{\mcitedefaultmidpunct}
{\mcitedefaultendpunct}{\mcitedefaultseppunct}\relax
\EndOfBibitem
\bibitem[Chen \latin{et~al.}(2012)Chen, Adato, and Altug]{Chen_AN2012b}
Chen,~K.; Adato,~R.; Altug,~H. \emph{ACS Nano} \textbf{2012}, \emph{6},
  7998--8006, PMID: 22920565\relax
\mciteBstWouldAddEndPuncttrue
\mciteSetBstMidEndSepPunct{\mcitedefaultmidpunct}
{\mcitedefaultendpunct}{\mcitedefaultseppunct}\relax
\EndOfBibitem
\bibitem[Bossard \latin{et~al.}(2014)Bossard, Lin, Yun, Liu, Werner, and
  Mayer]{Bossard_AN2014a}
Bossard,~J.~A.; Lin,~L.; Yun,~S.; Liu,~L.; Werner,~D.~H.; Mayer,~T.~S.
  \emph{ACS Nano} \textbf{2014}, \emph{8}, 1517--1524\relax
\mciteBstWouldAddEndPuncttrue
\mciteSetBstMidEndSepPunct{\mcitedefaultmidpunct}
{\mcitedefaultendpunct}{\mcitedefaultseppunct}\relax
\EndOfBibitem
\bibitem[Rozin \latin{et~al.}(2015)Rozin, Rosen, Dill, and Tao]{Rozin_NC2015a}
Rozin,~M.~J.; Rosen,~D.~A.; Dill,~T.~J.; Tao,~A.~R. \emph{Nature
  Communications} \textbf{2015}, \emph{6}, 7325 EP --\relax
\mciteBstWouldAddEndPuncttrue
\mciteSetBstMidEndSepPunct{\mcitedefaultmidpunct}
{\mcitedefaultendpunct}{\mcitedefaultseppunct}\relax
\EndOfBibitem
\bibitem[H{\"a}gglund \latin{et~al.}(2013)H{\"a}gglund, Zeltzer, Ruiz, Thomann,
  Lee, Brongersma, and Bent]{Hagglund_NL2013a}
H{\"a}gglund,~C.; Zeltzer,~G.; Ruiz,~R.; Thomann,~I.; Lee,~H.-B.-R.;
  Brongersma,~M.~L.; Bent,~S.~F. \emph{Nano Letters} \textbf{2013}, \emph{13},
  3352--3357, PMID: 23805835\relax
\mciteBstWouldAddEndPuncttrue
\mciteSetBstMidEndSepPunct{\mcitedefaultmidpunct}
{\mcitedefaultendpunct}{\mcitedefaultseppunct}\relax
\EndOfBibitem
\bibitem[Akselrod \latin{et~al.}(2015)Akselrod, Huang, Hoang, Bowen, Su, Smith,
  and Mikkelsen]{Akselrod_AM2015a}
Akselrod,~G.~M.; Huang,~J.; Hoang,~T.~B.; Bowen,~P.~T.; Su,~L.; Smith,~D.~R.;
  Mikkelsen,~M.~H. \emph{Advanced Materials} \textbf{2015}, \emph{27},
  8028--8034\relax
\mciteBstWouldAddEndPuncttrue
\mciteSetBstMidEndSepPunct{\mcitedefaultmidpunct}
{\mcitedefaultendpunct}{\mcitedefaultseppunct}\relax
\EndOfBibitem
\bibitem[Hedayati \latin{et~al.}(2011)Hedayati, Javaherirahim, Mozooni,
  Abdelaziz, Tavassolizadeh, Chakravadhanula, Zaporojtchenko, Strunkus, Faupel,
  and Elbahri]{Hedayati_AM2011a}
Hedayati,~M.~K.; Javaherirahim,~M.; Mozooni,~B.; Abdelaziz,~R.;
  Tavassolizadeh,~A.; Chakravadhanula,~V. S.~K.; Zaporojtchenko,~V.;
  Strunkus,~T.; Faupel,~F.; Elbahri,~M. \emph{Advanced Materials}
  \textbf{2011}, \emph{23}, 5410--5414\relax
\mciteBstWouldAddEndPuncttrue
\mciteSetBstMidEndSepPunct{\mcitedefaultmidpunct}
{\mcitedefaultendpunct}{\mcitedefaultseppunct}\relax
\EndOfBibitem
\bibitem[Aydin \latin{et~al.}(2011)Aydin, Ferry, Briggs, and
  Atwater]{Aydin_NC2011a}
Aydin,~K.; Ferry,~V.~E.; Briggs,~R.~M.; Atwater,~H.~A. \emph{Nat Commun}
  \textbf{2011}, \emph{2}, 517\relax
\mciteBstWouldAddEndPuncttrue
\mciteSetBstMidEndSepPunct{\mcitedefaultmidpunct}
{\mcitedefaultendpunct}{\mcitedefaultseppunct}\relax
\EndOfBibitem
\bibitem[Liu \latin{et~al.}(2015)Liu, Liu, Huang, Pan, Chen, Liu, and
  Gu]{Liu_AAMI2015a}
Liu,~Z.; Liu,~X.; Huang,~S.; Pan,~P.; Chen,~J.; Liu,~G.; Gu,~G. \emph{ACS
  Applied Materials \& Interfaces} \textbf{2015}, \emph{7}, 4962--4968, PMID:
  25679790\relax
\mciteBstWouldAddEndPuncttrue
\mciteSetBstMidEndSepPunct{\mcitedefaultmidpunct}
{\mcitedefaultendpunct}{\mcitedefaultseppunct}\relax
\EndOfBibitem
\bibitem[Ng \latin{et~al.}(2016)Ng, Cadusch, Dligatch, Roberts, Davis,
  Mulvaney, and Gomez]{Ng_AN0a}
Ng,~C.; Cadusch,~J.; Dligatch,~S.; Roberts,~A.; Davis,~T.~J.; Mulvaney,~P.;
  Gomez,~D.~E. \emph{ACS Nano} \textbf{2016}, \emph{10}, 4704--4711\relax
\mciteBstWouldAddEndPuncttrue
\mciteSetBstMidEndSepPunct{\mcitedefaultmidpunct}
{\mcitedefaultendpunct}{\mcitedefaultseppunct}\relax
\EndOfBibitem
\bibitem[Thongrattanasiri \latin{et~al.}(2012)Thongrattanasiri, Koppens, and
  Garc\'{\i}a~de Abajo]{Thongrattanasiri_PRL2012a}
Thongrattanasiri,~S.; Koppens,~F. H.~L.; Garc\'{\i}a~de Abajo,~F.~J.
  \emph{Phys. Rev. Lett.} \textbf{2012}, \emph{108}, 047401\relax
\mciteBstWouldAddEndPuncttrue
\mciteSetBstMidEndSepPunct{\mcitedefaultmidpunct}
{\mcitedefaultendpunct}{\mcitedefaultseppunct}\relax
\EndOfBibitem
\bibitem[Moreau \latin{et~al.}(2012)Moreau, Ciraci, Mock, Hill, Wang, Wiley,
  Chilkoti, and Smith]{Moreau_N2012a}
Moreau,~A.; Ciraci,~C.; Mock,~J.~J.; Hill,~R.~T.; Wang,~Q.; Wiley,~B.~J.;
  Chilkoti,~A.; Smith,~D.~R. \emph{Nature} \textbf{2012}, \emph{492},
  86--89\relax
\mciteBstWouldAddEndPuncttrue
\mciteSetBstMidEndSepPunct{\mcitedefaultmidpunct}
{\mcitedefaultendpunct}{\mcitedefaultseppunct}\relax
\EndOfBibitem
\bibitem[Lassiter \latin{et~al.}(2013)Lassiter, McGuire, Mock, Cirac{\`\i},
  Hill, Wiley, Chilkoti, and Smith]{Lassiter_NL2013a}
Lassiter,~J.~B.; McGuire,~F.; Mock,~J.~J.; Cirac{\`\i},~C.; Hill,~R.~T.;
  Wiley,~B.~J.; Chilkoti,~A.; Smith,~D.~R. \emph{Nano Letters} \textbf{2013},
  \emph{13}, 5866--5872, PMID: 24199752\relax
\mciteBstWouldAddEndPuncttrue
\mciteSetBstMidEndSepPunct{\mcitedefaultmidpunct}
{\mcitedefaultendpunct}{\mcitedefaultseppunct}\relax
\EndOfBibitem
\bibitem[Ng \latin{et~al.}(2012)Ng, Udagedara, Rukhlenko, Chen, Tang,
  Premaratne, and Cheng]{Ng_AN2012a}
Ng,~K.~C.; Udagedara,~I.~B.; Rukhlenko,~I.~D.; Chen,~Y.; Tang,~Y.;
  Premaratne,~M.; Cheng,~W. \emph{ACS Nano} \textbf{2012}, \emph{6}, 925--934,
  PMID: 22176669\relax
\mciteBstWouldAddEndPuncttrue
\mciteSetBstMidEndSepPunct{\mcitedefaultmidpunct}
{\mcitedefaultendpunct}{\mcitedefaultseppunct}\relax
\EndOfBibitem
\bibitem[Guo \latin{et~al.}(2014)Guo, Sikdar, Huang, Si, Su, Chen, Xiong, Yap,
  Premaratne, and Cheng]{Guo_TJOPCC2014a}
Guo,~P.; Sikdar,~D.; Huang,~X.; Si,~K.~J.; Su,~B.; Chen,~Y.; Xiong,~W.;
  Yap,~L.~W.; Premaratne,~M.; Cheng,~W. \emph{The Journal of Physical Chemistry
  C} \textbf{2014}, \emph{118}, 26816--26824\relax
\mciteBstWouldAddEndPuncttrue
\mciteSetBstMidEndSepPunct{\mcitedefaultmidpunct}
{\mcitedefaultendpunct}{\mcitedefaultseppunct}\relax
\EndOfBibitem
\bibitem[Si \latin{et~al.}(2014)Si, Sikdar, Chen, Eftekhari, Xu, Tang, Xiong,
  Guo, Zhang, Lu, Bao, Zhu, Premaratne, and Cheng]{Si_AN2014a}
Si,~K.~J.; Sikdar,~D.; Chen,~Y.; Eftekhari,~F.; Xu,~Z.; Tang,~Y.; Xiong,~W.;
  Guo,~P.; Zhang,~S.; Lu,~Y.; Bao,~Q.; Zhu,~W.; Premaratne,~M.; Cheng,~W.
  \emph{ACS Nano} \textbf{2014}, \emph{8}, 11086--11093, PMID: 25265019\relax
\mciteBstWouldAddEndPuncttrue
\mciteSetBstMidEndSepPunct{\mcitedefaultmidpunct}
{\mcitedefaultendpunct}{\mcitedefaultseppunct}\relax
\EndOfBibitem
\bibitem[Si \latin{et~al.}(2015)Si, Guo, Shi, and Cheng]{Si_AC2015a}
Si,~K.~J.; Guo,~P.; Shi,~Q.; Cheng,~W. \emph{Analytical Chemistry}
  \textbf{2015}, \emph{87}, 5263--5269, PMID: 25860874\relax
\mciteBstWouldAddEndPuncttrue
\mciteSetBstMidEndSepPunct{\mcitedefaultmidpunct}
{\mcitedefaultendpunct}{\mcitedefaultseppunct}\relax
\EndOfBibitem
\bibitem[Shi \latin{et~al.}(2018)Shi, Dong, Si, Sikdar, Yap, Premaratne, and
  Cheng]{Shi_AN2018a}
Shi,~Q.; Dong,~D.; Si,~K.~J.; Sikdar,~D.; Yap,~L.~W.; Premaratne,~M.; Cheng,~W.
  \emph{ACS Nano} \textbf{2018}, \emph{12}, 1014--1022, PMID: 29303252\relax
\mciteBstWouldAddEndPuncttrue
\mciteSetBstMidEndSepPunct{\mcitedefaultmidpunct}
{\mcitedefaultendpunct}{\mcitedefaultseppunct}\relax
\EndOfBibitem
\bibitem[Shi \latin{et~al.}(2016)Shi, Si, Sikdar, Yap, Premaratne, and
  Cheng]{Shi_AN2016a}
Shi,~Q.; Si,~K.~J.; Sikdar,~D.; Yap,~L.~W.; Premaratne,~M.; Cheng,~W. \emph{ACS
  Nano} \textbf{2016}, \emph{10}, 967--976, PMID: 26731313\relax
\mciteBstWouldAddEndPuncttrue
\mciteSetBstMidEndSepPunct{\mcitedefaultmidpunct}
{\mcitedefaultendpunct}{\mcitedefaultseppunct}\relax
\EndOfBibitem
\bibitem[Davis and G\'omez(2017)Davis, and G\'omez]{Davis_ROMP2016a}
Davis,~T.~J.; G\'omez,~D.~E. \emph{Reviews of Modern Physics} \textbf{2017},
  \emph{89}, 011003\relax
\mciteBstWouldAddEndPuncttrue
\mciteSetBstMidEndSepPunct{\mcitedefaultmidpunct}
{\mcitedefaultendpunct}{\mcitedefaultseppunct}\relax
\EndOfBibitem
\bibitem[Bauer \latin{et~al.}(2011)Bauer, Kobiela, and Giessen]{Bauer_PRB2011a}
Bauer,~C.; Kobiela,~G.; Giessen,~H. \emph{Phys. Rev. B} \textbf{2011},
  \emph{84}, 193104\relax
\mciteBstWouldAddEndPuncttrue
\mciteSetBstMidEndSepPunct{\mcitedefaultmidpunct}
{\mcitedefaultendpunct}{\mcitedefaultseppunct}\relax
\EndOfBibitem
\bibitem[Sherry \latin{et~al.}(2005)Sherry, Chang, Schatz, Van~Duyne, Wiley,
  and Xia]{Sherry_NL2005a}
Sherry,~L.~J.; Chang,~S.-H.; Schatz,~G.~C.; Van~Duyne,~R.~P.; Wiley,~B.~J.;
  Xia,~Y. \emph{Nano Letters} \textbf{2005}, \emph{5}, 2034--2038, PMID:
  16218733\relax
\mciteBstWouldAddEndPuncttrue
\mciteSetBstMidEndSepPunct{\mcitedefaultmidpunct}
{\mcitedefaultendpunct}{\mcitedefaultseppunct}\relax
\EndOfBibitem
\bibitem[Zhang \latin{et~al.}(2011)Zhang, Bao, Halas, Xu, and
  Nordlander]{Zhang_NL2011a}
Zhang,~S.; Bao,~K.; Halas,~N.~J.; Xu,~H.; Nordlander,~P. \emph{Nano Letters}
  \textbf{2011}, \emph{11}, 1657--1663\relax
\mciteBstWouldAddEndPuncttrue
\mciteSetBstMidEndSepPunct{\mcitedefaultmidpunct}
{\mcitedefaultendpunct}{\mcitedefaultseppunct}\relax
\EndOfBibitem
\bibitem[Prodan \latin{et~al.}(2003)Prodan, Radloff, Halas, and
  Nordlander]{Prodan_S2003a}
Prodan,~E.; Radloff,~C.; Halas,~N.~J.; Nordlander,~P. \emph{Science}
  \textbf{2003}, \emph{302}, 419--422\relax
\mciteBstWouldAddEndPuncttrue
\mciteSetBstMidEndSepPunct{\mcitedefaultmidpunct}
{\mcitedefaultendpunct}{\mcitedefaultseppunct}\relax
\EndOfBibitem
\bibitem[Cirac{\`\i} \latin{et~al.}(2013)Cirac{\`\i}, Lassiter, Moreau, and
  Smith]{Ciraci_JOAP2013a}
Cirac{\`\i},~C.; Lassiter,~J.~B.; Moreau,~A.; Smith,~D.~R. \emph{Journal of
  Applied Physics} \textbf{2013}, \emph{114}, 163108\relax
\mciteBstWouldAddEndPuncttrue
\mciteSetBstMidEndSepPunct{\mcitedefaultmidpunct}
{\mcitedefaultendpunct}{\mcitedefaultseppunct}\relax
\EndOfBibitem
\bibitem[Bowen and Smith(2014)Bowen, and Smith]{Bowen_PRB2014a}
Bowen,~P.~T.; Smith,~D.~R. \emph{Phys. Rev. B} \textbf{2014}, \emph{90},
  195402\relax
\mciteBstWouldAddEndPuncttrue
\mciteSetBstMidEndSepPunct{\mcitedefaultmidpunct}
{\mcitedefaultendpunct}{\mcitedefaultseppunct}\relax
\EndOfBibitem
\bibitem[Chen(2012)]{Chen_OE2012a}
Chen,~H.-T. \emph{Opt. Express} \textbf{2012}, \emph{20}, 7165--7172\relax
\mciteBstWouldAddEndPuncttrue
\mciteSetBstMidEndSepPunct{\mcitedefaultmidpunct}
{\mcitedefaultendpunct}{\mcitedefaultseppunct}\relax
\EndOfBibitem
\bibitem[Qianqian \latin{et~al.}(2018)Qianqian, Debabrata, Runfang, Jye,
  Dashen, Yiyi, Malin, and Wenlong]{Qianqian_AMa}
Qianqian,~S.; Debabrata,~S.; Runfang,~F.; Jye,~S.~K.; Dashen,~D.; Yiyi,~L.;
  Malin,~P.; Wenlong,~C. \emph{Advanced Materials} \textbf{2018}, 1801118\relax
\mciteBstWouldAddEndPuncttrue
\mciteSetBstMidEndSepPunct{\mcitedefaultmidpunct}
{\mcitedefaultendpunct}{\mcitedefaultseppunct}\relax
\EndOfBibitem
\bibitem[Johnson and Christy(1972)Johnson, and Christy]{Johnson_PRB1972a}
Johnson,~P.~B.; Christy,~R.~W. \emph{Phys. Rev. B} \textbf{1972}, \emph{6},
  4370--4379\relax
\mciteBstWouldAddEndPuncttrue
\mciteSetBstMidEndSepPunct{\mcitedefaultmidpunct}
{\mcitedefaultendpunct}{\mcitedefaultseppunct}\relax
\EndOfBibitem
\bibitem[Raki\'c \latin{et~al.}(1998)Raki\'c, Djuri\v{s}ic, Elazar, and
  Majewski]{Rakic}
Raki\'c,~A.~D.; Djuri\v{s}ic,~A.~B.; Elazar,~J.~M.; Majewski,~M.~L.
  \emph{Applied Optics} \textbf{1998}, \emph{37}, 5271--5283\relax
\mciteBstWouldAddEndPuncttrue
\mciteSetBstMidEndSepPunct{\mcitedefaultmidpunct}
{\mcitedefaultendpunct}{\mcitedefaultseppunct}\relax
\EndOfBibitem
\end{mcitethebibliography}

\setcounter{equation}{0}
\renewcommand{\theequation}{S.\arabic{equation}}
\setcounter{figure}{0}
\renewcommand{\thefigure}{S.\arabic{figure}}
\appendix
\newpage

\begin{figure}
	\centering
	\includegraphics[width=80mm]{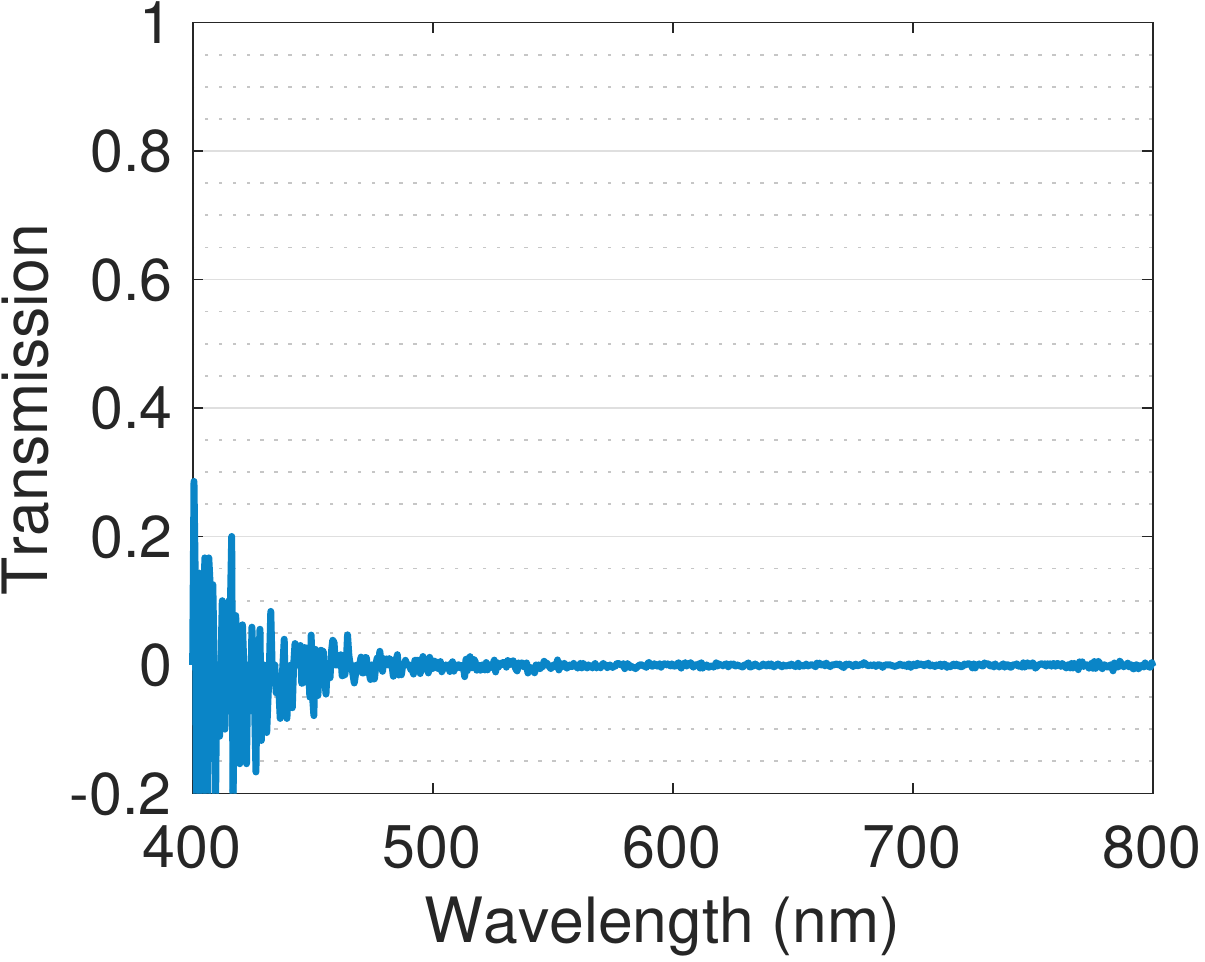}\\
    \includegraphics[width=80mm]{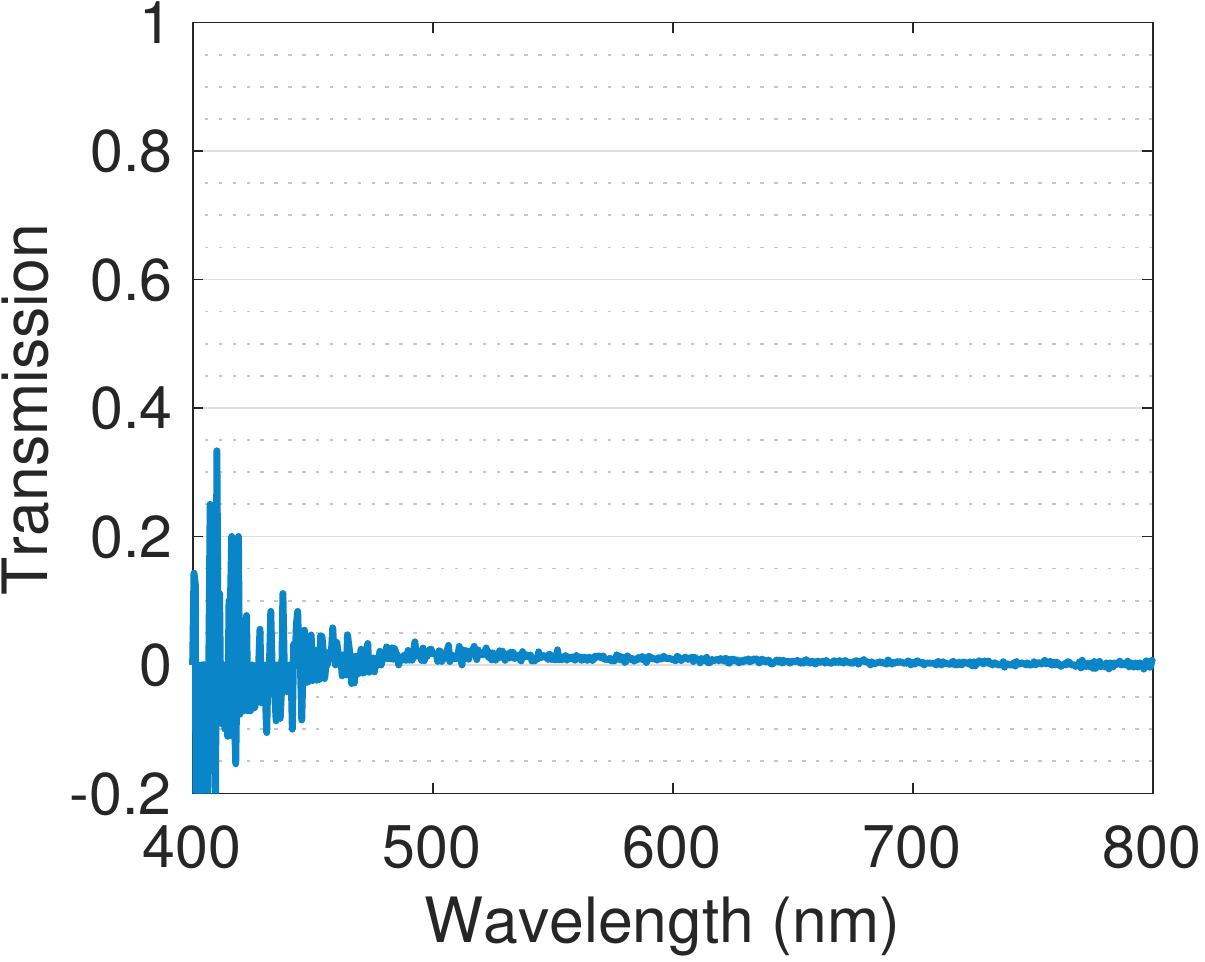}
	\caption{Measured transmission spectra of for the plasmene absorbers corresponding to Figure 3 of the main text, for the case of an (top) Al and a (bottom) Au mirror.}
	\label{fig:s2}
\end{figure}

\begin{figure}
	\centering
	\includegraphics[width=80mm]{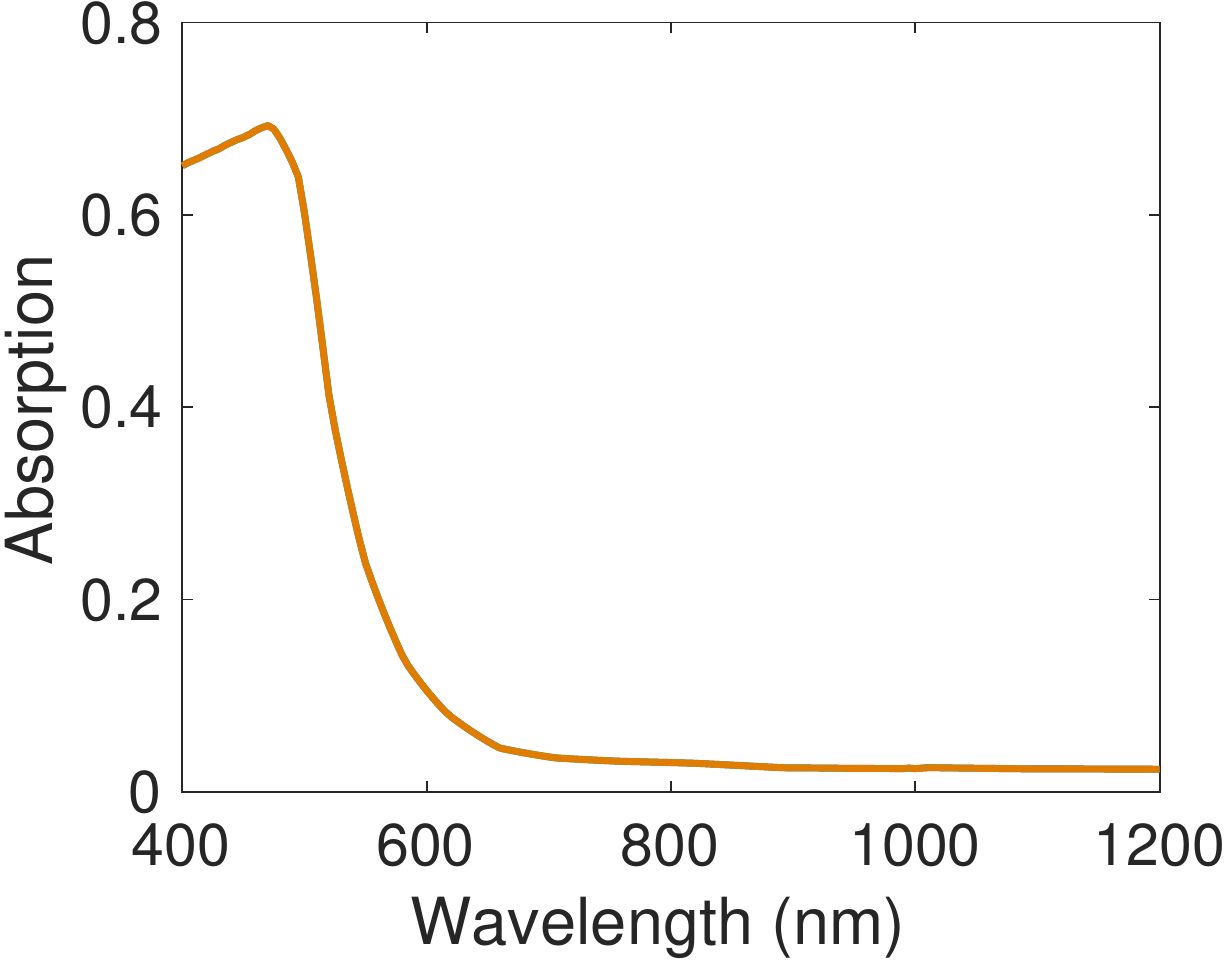}
	\caption{Calculated absorption spectrum of a 40 nm thick Au film placed on top of a 30 nm thick TiO$_2$ / Au mirror support.}
	\label{fig:s1}
\end{figure}

\end{document}